\documentclass[final]{svjour2}
\usepackage{graphicx}
\usepackage{rotating}
\usepackage{amssymb}
\usepackage{mathptmx}
\usepackage[numbers]{natbib}
\makeatletter
\journalname{Journal of Low Temperature Physics}
\newcommand{\hefour}{$^4$He}

\bibpunct{}{}{,}{s}{}{,}

\begin{document}

\newcommand{\hdblarrow}{H\makebox[0.9ex][l]{$\downdownarrows$}-}
\title{ The glassy response of double torsion oscillators in solid $^4$He}

\author{Matthias J. Graf$^1$ \and Jung-Jung Su$^{1,2}$ \and Hari P. Dahal$^{1,2}$ 
\and I. Grigorenko$^{1,3}$ \and Zohar Nussinov$^4$}

\institute{1: 
Theoretical Division, Los Alamos National Laboratory, Los Alamos, New Mexico 87545, USA
\\2: 
Center for Integrated Nanotechnologies, Los Alamos National Laboratory, Los Alamos, New Mexico 87545, USA
\\ 3:
Department of Physics, Penn State University, University Park, Pennsylvania 16802, USA
\\ 4:
Department of Physics, Washington University, St. Louis, Missouri 63160, USA
\\
\email{graf@lanl.gov}
}

%\date{20.06.2010}
\date{30.11.2010}

\maketitle

\begin{abstract}
Single and double torsion oscillators have been used to measure the anomalous change in resonant frequency and 
accompanying dissipation  in solid $^4$He.
We present a glass description of the mechanical anomalies found in torsion oscillator measurements.
Our results show that it is not necessary to invoke a supersolid interpretation to explain these mechanical anomalies. 
Previously, we demonstrated that the back-action of a glassy subsystem present in solid $^4$He can account for 
frequency change and dissipation peak in single torsion oscillator experiments.
Here, we show that the same glassy back-action can explain the experimental results of the composite
torsion oscillator developed by the Rutgers group, which measures the response of
solid \hefour\ at the in-phase mode $f_1=496$ Hz and out-of-phase mode $f_2=1173$ Hz.
%PACS numbers: 67.80.B-, 67.80.bd, 64.70.P-
% 67.80.B- solid phase, He-4
% 67.80.bd supersolid, He-4
% 64.70.P- Glass transitions
\end{abstract}


\section{Introduction}

The controversy over the origin of the anomalous low-temperature signature in solid \hefour\ continues
since it was first discovered in ultrasound measurements by Goodkind and collaborators\cite{Lengua, Ho, Goodkind} 
and eventually detected in torsion oscillator (TO) experiments by Kim and Chan.\cite{Kim04, Kim05}
In earlier work,
we argued that the origin of the TO anomaly is not 
that of a non-classical rotational inertia effect 
(NCRI), but rather due to {\em a mechanical effect}.
We examined the response of
a nonuniform solid with defects leading to transient elastic dynamics at high
temperatures.\cite{Nussinov07,Graf09,Graf10,Su10b,Su10c}
These defects are quenched at lower 
temperatures leading to increased mechanical stiffness 
and hence increased rotational speed of the TO. Furthermore, a nonuniform
solid with a small concentration of defects
can explain the maximal damping at the crossover temperature where
the putative NCRI was reported.  Our defect-based scenario
is quite simple. At high temperatures, the transient dynamics is 
very rapid relative to the time scale set by the TO, while at temperatures far lower than the putative supersolid
transition, the transient dynamics is much longer.
As in classical damped oscillators, when the transient dynamics 
due to internal dissipation matches the impedance of the external driving force, the dissipation peaks.

A ``freezing'' between a high-temperature dynamic state and a
low-temperature immobile state may be characterized by either (i) a discontinuous first order or
continuous transition
in annealed systems with no disorder or (ii) an observed dynamic crossover in amorphous 
systems with disorder (spin-glass) or self-generated 
quenched disorder due to 
rapid cooling (as in structural glasses).  
Freezing of rapid degrees of freedom can be triggered by a distribution of activation barriers;
as we will briefly review below,  activated dynamics corresponds to a particular choice of the relaxation time for glasses. Such a distribution
of barriers also lies at the heart of theories describing insulating ``electron glasses'' in which electrons are pinned by disorder. \cite{electron_glass}
Some theories of structural glasses assume that there is no true glass ``transition'', but rather a dynamic crossover
at which the relaxation time becomes long relative to
experimental time scales. Yet other theories assume that
an ideal glass exists in the thermodynamic limit but that, experimentally, 
it is largely inaccessible by divergent equilibration times. Notwithstanding the lack of consensus
on the origin of universal slow dynamics in numerous amorphous systems, the increase in observed relaxation 
times is due to pinning or freezing of some form of degrees of freedom that are dynamic at high temperature. \cite{bert}

Several theories of glasses assume that elastic defects can entangle
and become pinned  and consequently form a sluggish state at low temperatures. \cite{nelsonGF}
An early theory of defect dynamics investigated the consequences of a low-temperature pinning of dislocations.\cite{granato56} 
Although, as emphasized above, the underlying mechanism of low temperature dynamics
in these amorphous systems is not universally agreed upon, {\em their empirical behavior
has precise universal features}, \cite{lubchenkowolynes}
which we employ here.
Salient features of amorphous systems that are
pertinent to our study include:

$\bullet$ A nearly universal Vogel-Fulcher-Tammann (VFT) form describing the increase 
in relaxation time of glasses as temperature is lowered. The VFT form
has been used to fit data collected from many glasses, although
it has no unique theoretical basis. Specifically, the VFT  expression
$\tau = \tau_0 \exp(\Delta/(T-T_{0}))$ includes activated (Arrhenius)
dynamics ($T_0=0$) dominated by energy barriers $\Delta$.
Thus, the general VFT analysis that we employ includes the specific case of 
such activated dynamics; this particular case was discussed in our earlier fits to the TO data. \cite{Nussinov07}
Indeed, activated dynamics triggered by pinning of defects (dislocations) may occur
in solid \hefour. \cite{Balatsky07} Similarly, increased
dynamics relative to an activated form ($T_{0}<0$)
suggests an increase in motion due to quantum fluctuations. 

$\bullet$ The activated dynamics in amorphous systems is characterized by
a broad {\em distribution of relaxation times}.  The empirical response functions, 
which capture this distribution, are consequently of the 
Davidson-Cole or Cole-Cole form which we invoke in our analysis. 

$\bullet$ Omni-present memory effects and hysteresis 
are particularly important in these systems and characterize
the observed dynamic crossover including ``aging'' 
behavior.\cite{memory1,memory2,memory3} 
Such memory effects are of current technological importance. \cite{memory4} 

$\bullet$ Finally, low-temperature specific heat contributions that scale linearly with temperature in insulators
have been key indicators for glassiness.\cite{Esquinazi,Enss}

It is important to emphasize that our approach assumes that only a small glass-like fraction present in solid \hefour\ 
exhibits a freezing transition of a high-temperature mobile component 
with a distribution of relaxation times. \cite{Nussinov07} 
Since we first proposed a nonsupersolid interpretation, several other disorder-based theories have been put
forward to explain the large dissipation seen in torsion oscillator or shear modulus
experiments.
Yoo and Dorsey proposed that solid \hefour\ displays viscoelastic properties.\cite{Yoo2009} 
Syshchenko et al.\cite{Syshchenko2009,Syshchenko2010} and Iwaza\cite{Iwasa2010} argued that the shear modulus measurements can
be understood in terms of dislocation motion and the formation of a low-temperature dislocation network  based
on the theory by Granato and L\"ucke.\cite{granato56}
All these models have in common disorder, but do not require glassiness.
Similarly, Syshchenko\cite{Syshchenko2010} speculated that the observed
dissipation peak is associated with boiling off of $^3$He atoms
from dislocation lines. The $^3$He unbinding causes a depinning with increasing temperature that generates a distribution
of mobile dislocation lines responsible for softening of the shear modulus and increased dissipation.
Very recently, Reppy\cite{Reppy2010} reported a nonsupersolid scenario in TO experiments,
where induced disorder affected primarily the high-temperature response above the putative
supersolid transition. A result consistent with the dynamics of highly disordered or glassy systems.
So far attempts to offer a unified picture within elasticity theory of uniform solids, 
i.e., how applied shear stresses affect torsion oscillators and elastic coefficient measurements,
have been unsuccessful.\cite{Syshchenko2009,Clark2008,Maris2010} 
This may be due to a neglect of anelastic contributions in these approaches.

%%
% Fig. 1
%%
\begin{figure}
%\bigskip
\bigskip
\begin{center}
\includegraphics[ width=0.9\linewidth, keepaspectratio, angle=0 ]{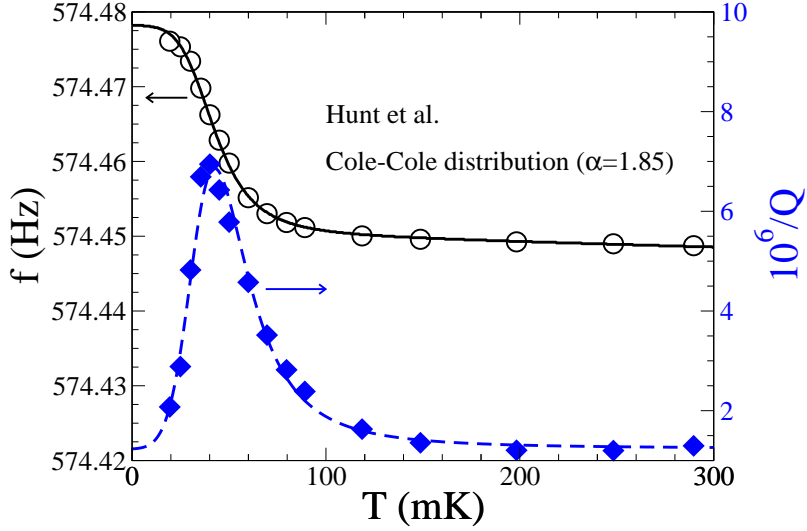}
\end{center}
\caption{(Color online) Torsion oscillator resonant frequency (black, left axis) and dissipation (blue, right axis) vs.\ temperature.
The experimental data (symbols) for a single torsion oscillator
are from Hunt et al.\cite{Hunt09} Details of the glass model (curves), the Cole-Cole distribution function of Debye 
relaxors with exponent $\alpha=1.85$, activation barrier $\Delta = 408$ mK, and other model parameters are given in Ref.~\cite{Graf10}
}\label{Fig1}
\end{figure}

Since a true phase transition is a bulk effect of the sample,
we investigated previously thermodynamic signatures as reported by specific heat and pressure
measurements.\cite{Lin07, Lin09, Grigorev07a, Grigorev07b, Rudavskii10}
Our thermodynamic analysis in terms of two-level systems is consistent with a glassy interpretation of the 
linear-$T$ dependence in the specific heat below $\sim 100$ mK 
and the quadratic-$T$ dependence in pressure measurements of the equation of state, $P(T)$,
of an otherwise perfect Debye solid.\cite{Balatsky07, Su10a}
A Debye solid exhibits a specific heat $C \propto T^3$ and a pressure dependence
$P \propto T^4$.
It is notable that the associated excess entropy is at least three orders of magnitude too small to explain an NCRI effect
of a couple of percent in the TO experiments if it is caused by uniform Bose-Einstein condensation.\cite{Balatsky07, Su10a}
Finally, we demonstrated previously that single torsion oscillators can be described by using a 
distribution of glassy relaxation times, for example, the well-known Cole-Cole distribution.
In Fig.~\ref{Fig1} we show excellent agreement between model calculations for a glassy back-action
and the data by Hunt and coworkers for a single TO.\cite{Hunt09} 
Details of the model have been described in Ref.~\cite{Graf10} 
The important message of this comparison is that 100\% of the TO signal can be ascribed to
a glassy back-action response of mechanical origin with no need for a supersolid component.

Here we focus on the TO experiments reported by Aoki and coworkers.\cite{Aoki07,Aoki08} 
The unique design of two coupled torsion oscillators has opened up a window in frequency space to explore
the dynamic response of solid \hefour\ at two very different frequencies. 
However, these measurements have been difficult to explain within a supersolid or glassy scenario alike.
In the past, these difficulties have been used as arguments against a glassy interpretation.\cite{Aoki07,Aoki08,Clark2008}
The composite design allowed for the measurement of an in-phase, $f_1=496$ Hz,
and out-of-phase, $f_2=1173$ Hz, resonant frequency on the same solid \hefour\ sample. 
Surprisingly, the observed changes in frequency and dissipation as a function of temperature showed that 
the relative magnitude in frequency shift, 
$\Delta f_i/f_i=(f_i(0\, {\rm K}) - f_i(0.3\, {\rm K}))/f_i(0.3\, {\rm K})$,
is nearly frequency independent.
At the same time, the dissipation peak shifted only slightly to higher temperatures at the higher frequency.
These observations are contrary to the standard theory of supersolidity, with the hallmark of dissipationless superflow
of vacancies or interstitials. Even when invoking a vortex picture for supersolidity to account for the anomalously
large dissipation, it fails to account for the unchanged relative shift in resonant frequency 
$\Delta f_1/f_1 \approx \Delta f_2/f_2$, which is a direct measure of the NCRI.
Any theory of superfluidity predicts that the superfluid fraction is more rapidly destroyed when perturbed at higher frequency.
This is certainly at odds with the experiments by Aoki and coworkers and deserves more attention.
Simply extrapolating results of the single TO to higher frequencies fails to explain the
experiments.  
The puzzle is how to explain the outcome of the composite double TO experiment.

In this paper, we present an interpretation of these measurements using a glassy back-action term in the 
response function of the double TO similar to our earlier work for the single TO.
We provide a consistent interpretation and modeling of the composite double TO within 
the glassy framework, where we postulate the presence of a glassy subsystem in solid \hefour. 
We descibe the experiments by Aoki in terms of two coupled  mechanical torsion oscillators,
shown schematically in Fig.~\ref{Fig2}.
Our model can explain the measurements, though requires anomalous damping and a strongly frequency dependent glass
term.
Note that anomalous damping is already required for the empty cell (no \hefour\ present!) to explain that the quality factor 
$Q_2$ of the high mode is lower than $Q_1$ of the low mode at  0.3 K,
its origin may be due to anelasticity or non-rigid connections between the torsion rod  
or to a nonlinear feedback of the capacitive drive onto the dummy bob.
Additionally, our results point toward \hefour\ slipping at the container walls of the cylindrical pressure cell consistent 
with the very small, frequency-independent relative shifts 
$\Delta f_1/f_1 \approx \Delta f_2/f_2 \approx 1.7\cdot 10^{-6}$.

\section{A model for the coupled double oscillator}

%%
% Fig. 2
%%
\begin{figure}
\bigskip
\begin{center}
\includegraphics[ width=0.30\linewidth, keepaspectratio, angle=0 ]{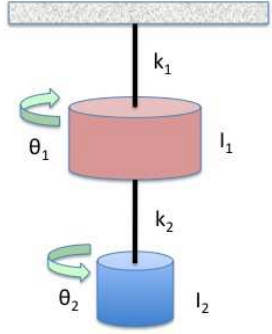}
\end{center}
\caption{(Color online) Sketch of the double torsion oscillator modeled in Eq.~(\ref{EOM}).
The upper moment of inertia ($I_1$) is the dummy bob, while the lower moment of inertia ($I_2$) 
is the cylindrical pressure cell that can be loaded with \hefour. The stiffness of the BeCu torsion rods is given by
$k_1$ and $k_2$ with $k_1 \approx k_2$ by design.}
\label{Fig2}
\end{figure}

We model the coupled double torsion oscillator of the Rutgers group, sketched in Fig.~\ref{Fig2}, 
by the following system of equations for the torsion angles:
\begin{eqnarray}\label{EOM}
  I_1 \ddot \Theta_1(t) + \gamma_1 \dot \Theta_1(t) + k_1 \Theta_1(t) + k_2 (\Theta_1(t) - \Theta_2(t)) &=& F(t) ,
\nonumber\\ 
  I_2 \ddot \Theta_2(t) + \gamma_2 \dot \Theta_2(t) + k_2 (\Theta_2(t) - \Theta_1(t)) &=&
  \int dt' g(t-t') \Theta_2(t') ,
\end{eqnarray}
where $\Theta_i(t)$ are torsion angles, $\gamma_i$ are damping coefficients, $k_i$ are torsion rod 
stiffnesses, $g(t)$ is the glass back-action term, and $F(t)$ is the applied external torque.
The subindex ``$1$'' refers to the upper or dummy bob in the experiment, 
while ``$2$'' refers to the lower oscillator with the pressure cell that can be loaded with solid \hefour.
After Fourier transformation of Eq.~(\ref{EOM}), we obtain
\begin{eqnarray}\label{eqns}
  \left( -I_1 \omega^2 - i \gamma_1 \omega + k_1 + k_2 \right) \Theta_1(\omega) - k_2 \Theta_2(\omega) &=& F(\omega) ,
\nonumber\\ 
  \left( -I_2 \omega^2 - i \gamma_2 \omega + k_2 - g(\omega) \right) \Theta_2(\omega) - k_2 \Theta_1(\omega) &=& 0 .
\end{eqnarray}
For a strongly underdamped oscillator and a small glassy back-action, it suffices to solve first for the
bare resonant frequencies and later include perturbatively damping and glass terms.
The bare resonant frequencies ($F=0$) are
\begin{equation}\label{bare_freq}
\left( \omega_{1/2}^0\right)^2 = 
\frac{k_2 I_1 + (k_1+k_2) I_2 \mp \sqrt{ k_2^2 I_1^2 + (k_1+k_2)^2 I_2^2 + 2 I_1 I_2 k_2 (k_2-k_1) }}{2 I_1 I_2} .
\end{equation}
Next, we expand the system of equations in (\ref{eqns}) around the bare resonant frequencies (\ref{bare_freq}) by inserting
$\omega_i = \omega_i^0 + \Delta \omega_i$ and solving the secular equations for
\begin{equation}
\Delta \omega_i = -\frac{ g_1(\omega_i^0) R_{1}(\omega_i^0) + i 
 \left[ g_2(\omega_i^0) R_1(\omega_i^0) + \omega_i^0 ( \gamma_1 R_2(\omega_i^0) + \gamma_2 R_1(\omega_i^0) ) 
 \right]}{ 2 \omega_i^0 [I_1 R_2(\omega_i^0) + I_2 R_1(\omega_i^0)]} ,
\end{equation}
where we introduced ancillary functions for compactness of notation
\begin{eqnarray}
g(\omega) &=& g_1(\omega) + i g_2(\omega) , \\
R_1(\omega) &=& -\omega^2 I_1 + k_1 + k_2 , \\
R_2(\omega) &=& -\omega^2 I_2 + k_2 .
\end{eqnarray}
It is now straightforward to calculate the resonant frequencies and dissipation of the loaded oscillator
\begin{eqnarray}
f_i &=& \frac{ {\rm Re}\ \omega_i}{2 \pi} ,	\\
Q_i^{-1} &=& -\frac{ 2 {\rm Im}\ \omega_i }{\omega_i^0} .
\end{eqnarray}
At high temperatures, where $\omega\tau \to 0$ and $g_2 \approx 0$ the dissipation is given by
\begin{equation}
Q_{i, \infty}^{-1} \approx 
 \frac{ \gamma_1  R_2(\omega_i^0) + \gamma_2 R_1(\omega_i^0) }{ \omega_i^0 
 [I_1 R_2(\omega_i^0) + I_2 R_1(\omega_i^0)]} .
\end{equation}
Since the empty pressure cell shows significant temperature dependence in its response, we follow the experimental
analysis and subtract the $T$-dependent empty cell response and correct for the differences due to the small difference 
in $I_2$.
Hence, we compare against corrected shifts relative to a high-temperature value, e.g., at 300 mK,
\begin{equation}
\Delta f_i = f_i(T) - f_i(0.3\,{\rm K}).
\end{equation}
Within our notation we find that approximately $Q_{i, \infty}^{-1} \approx Q_i^{-1}(0.3\,{\rm K})$.

%%
% Fig. 3
%%
\begin{figure}
\begin{center}
\includegraphics[ width=1.0\linewidth, keepaspectratio, angle=0 ]{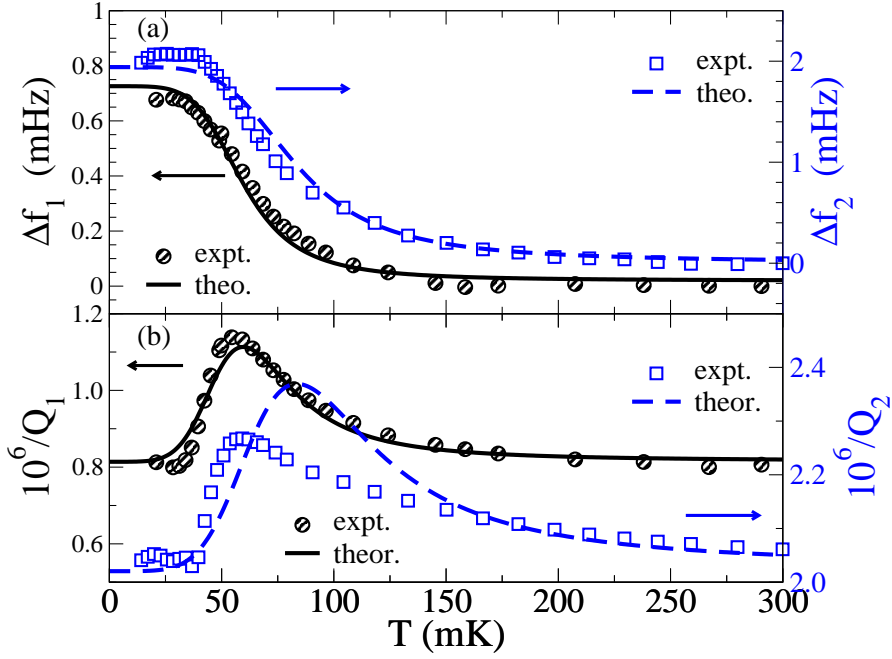}
\end{center}
\caption{(Color online) Frequency and dissipation in double torsion oscillator by Aoki et al.\cite{Aoki07}
(symbols) compared with glass theory (lines).
Panel (a): Temperature dependence of resonant frequency shifts $\Delta f_1$ (black, left axis) and $\Delta f_2$ (blue, right axis).
Panel (b): Temperature dependence of dissipation $Q_1^{-1}$ (black, left axis) and $Q_2^{-1}$ (blue, right axis).
The experimental data were corrected for the strongly temperature-dependent background of the empty cell.~\cite{Aoki07}
}
\label{fig_DO}
\end{figure}

Our model of the coupled composite oscillators can describe the temperature behavior of the Rutgers experiment if  we
make further assumptions about the glassy back-action $g(\omega)$. 
We model the dynamics of the glassy subsystem by
\begin{equation}
g(\omega) = \frac{ {\cal G}(\omega) }{ 1 - (i \omega \tau)^\alpha } .
\end{equation}
Here the glass term is ${\cal G}(\omega) = g_0 \left(\frac{\omega}{\omega_1^0}\right)^p$
and the Vogel-Fulcher-Tammann relaxation time is
$\tau(T) = \tau_{0} \exp[ \Delta/(T-T_{0})]$ for $T>T_{0}$.
$T_{0}$ is the ideal glass transition temperature, which is below
the temperature where the peak in dissipation occurs. The parameter $\Delta$ is an 
average potential barrier of the glass.
Finally, at temperatures $T<T_{0}$ the glassy subsystem freezes out and $\tau$ becomes infinite.

The parameters $I_i$ and $k_i$ can be determined from the bare resonant frequencies $f_i^0 = \omega_i^0/2\pi$
in Eq.~(\ref{bare_freq}).
In addition, the damping coefficients $\gamma_i$ can be extracted from the high-temperature dissipation $Q_{i, \infty}^{-1}$.
Finally, the glassy back-action ${\cal G}(\omega)$ with its parameterization accounts through $\tau(T)$
for the temperature dependence of 
$\Delta f_i$ and $Q_i^{-1}$.
Our phenomenological glass theory, applied to the coupled double oscillator, explains both
frequency shift and dissipation peak for in-phase and out-of-phase
torsional response.\cite{Aoki07}
Data for in-phase frequency $f_1=496$ Hz and out-of-phase $f_2=1173$ Hz 
are shown in Fig.~\ref{fig_DO}, plotted against the
temperature. The data are found to be well described by our model of coupled oscillators presented in Eq.~(\ref{eqns})
with model parameters:
$I_1 = 620.3$ nNs$^2$, 
$I_2 = 233.2$ nNs$^2$, 
$k_1 = 9.125$ N,
$k_2 = 8.355$ N,
$\gamma_1 = -16.93$ nNs,
$\gamma_2 = 10.43$ nNs,
$g_0 = 15.60$ $\mu$N,
$\tau_0 = 4.906$ $\mu$s,
$\Delta = 386.2$ mK,
$T_0 = -32.73$ mK,
$\alpha = 1.74$,
$p = 1.77$.
The obtained values for moment of inertia and rod stiffness agree well with
estimates for moments of inertia and BeCu rods.\cite{Aoki08}
At the highest measured temperatures, where the glassy contribution is small,
these parameters result in resonant frequencies 
$f_1(0.291\,{\rm K})=495.82813$ Hz, 
$f_2(0.300\,{\rm K}) = 1172.8159$ Hz 
and dissipation 
$Q_{1}^{-1}(0.291\,{\rm K}) = 0.820\cdot 10^{-6}$, 
$Q_{2}^{-1}(0.300\,{\rm K}) = 2.051\cdot 10^{-6}$ 
in excellent agreement with experiment. The experimental values are
$f_1^{expt.}(0.291\,{\rm K})= 495.82811$ Hz,
$f_2^{expt.}(0.300\,{\rm K}) = 1172.8158$ Hz 
and
$(Q_{1}^{-1})^{expt.}(0.291\,{\rm K}) = 0.806\cdot 10^{-6}$,
$(Q_{2}^{-1})^{expt.}(0.300\,{\rm K}) = 2.061\cdot 10^{-6}$.

It is worth pointing out that an anomalous damping coefficient $\gamma_1 \sim -\gamma_2$
is needed to explain the anomalous behavior of increased
dissipation with increased frequency. Note that this anomalous damping
is already required to describe the unloaded pressure cell, so it is unrelated to the
properties of solid \hefour.
After loading the cell with solid \hefour\ the dissipation ratio becomes 
$Q_{2}^{-1}/Q_{1}^{-1} = 2.5$ at 300 mK with frequency ratio $f_2/f_1 = 2.37$.
A negative $T_0$ is indicative of strong quantum fluctuations leading to an avoided
glass transition. This behavior resembles that of a paramagnetic system with a negative Curie-Weiss
temperature in the presence of antiferromagnetic fluctuations. It will be interesting to
see if the zero-point motion of \hefour\ is indeed responsible for this behavior.

Finally, the comparison in Fig.~\ref{fig_DO} shows that an explicit 
frequency-dependent back-action  must be used with 
${\cal G}(\omega) = g_0 \left(\frac{\omega}{\omega_1^0}\right)^p$
and $p = 1.77$ to account for the experimental fact of $\Delta f_1/f_1 \approx \Delta f_2/f_2$,
i.e., the relative frequency shift remains unchainged with changing resonant frequency.
In the past, when we studied single oscillators, it sufficed to parameterize ${\cal G}(\omega)$ by the
single resonant frequency, namely, ${\cal G}(\omega)\approx {\cal G}(\omega_i^0) = g_0$.

It will require further studies to sort out
whether the negative damping is related to an anelastic or non-rigid torsion rod or nonlinear feedback from the external drive.
Notice that the combined oscillator system is underdamped and dissipates energy with $Q_i > 0$.
On the other side, a frequency exponent of $p = 1.77$ in the glass term ${\cal G}(\omega)$ points toward slip of solid 
\hefour\ at the cylindrical container wall. 
Theories describing solid \hefour\ in torsion oscillators as a viscoelastic material \cite{Yoo2009} 
or two-level systems moving through a solid matrix \cite{Andreev07, Andreev09, Korshunov09}
predict an explicit frequency exponent of $p=4$ for the back-action term, when
assuming a no-slip boundary condition at the cylindrical container wall. So it is reasonable to expect
that the slip of \hefour\ at the container
wall will lead to a reduction in the power of the frequency dependence of ${\cal G}(\omega)$.

\section{Conclusions}
For the first time, we present a quantitative explanation of the anomalous frequency and dissipation 
dependence reported for the double torsion oscillator 
with cylindrical sample chamber by the Rutgers group
invoking glass dynamics.
So far these experiments have been at odds with supersolid and glassy interpretations.
The conventional single oscillator models fail to account simultaneously for the low and high resonant frequency
data.
Our studies of the coupled oscillator show that the observed shifts in resonant frequencies and dissipation are in agreement 
with a glassy back-action
contribution in solid \hefour\ if one includes anomalous damping in the dummy bob and explicit frequency dependence 
of the glassy back-action term.
Surprisingly, already the double torsion oscillator with the empty pressure cell requires a negative
damping coefficient for the dummy bob to accurately describe frequencies and dissipation at 300 mK.
Clearly, more dynamic studies of solid \hefour\ with coupled oscillators are needed to determine the frequency dependence over
a larger frequency range and for different cell designs to identify the origin of the anomalous damping and the frequency dependence of the back-action term.
It remains to be seen if a quantitative supersolid interpretation of the double oscillator experiment is possible.

\begin{acknowledgements}
We are grateful to H. Kojima and Y. Aoki for explaining their experiments and sharing their data. 
We like to thank A.V. Balatsky, J.C. Davis, J.M. Goodkind, and S.E. Korshunov for many stimulating discussions.
This work was supported by LDRD through the US Dept.\ of Energy at Los Alamos National Laboratory
under contract No.~DE-AC52-06NA25396 and by the Center for Materials
Innovation (CMI) of Washington University, St.\ Louis.

\end{acknowledgements}


\begin{thebibliography}{10}
\bibitem{Lengua} G. Lengua and J. M. Goodkind, J. Low Temp. Phys. {\bf 79}, 251 (1990).
\bibitem{Ho} P.-C. Ho, I. P. Bindloss, and J. M. Goodkind, J. Low Temp. Phys. {\bf 109}, 409 (1997).
\bibitem{Goodkind} J. M. Goodkind, Phys. Rev. Lett. {\bf 89}, 095301 (2002).

\bibitem{Kim04} E. Kim and M. H. W. Chan, Nature (London) {\bf 427}, 225 (2004).
\bibitem{Kim05} E. Kim and M. H. W. Chan, Science {\bf 305}, 1941 (2005).

\bibitem{Nussinov07} Z. Nussinov, M. J. Graf, A. V. Balatsky, and S. A. Trugman, Phys. Rev. B {\bf 76}, 014530 (2007).

\bibitem{Graf09} M. J. Graf, A. V. Balatsky, Z. Nussinov, I. Grigorenko, and S. A. Trugman, JPCS {\bf 150}, 032025 (2009).

\bibitem{Graf10} M. J. Graf, Z. Nussinov, and A. V. Balatsky, J. Low Temp. Phys.  {\bf 158}, 550 (2010).


\bibitem{Su10b} J.-J. Su, M. J. Graf, and A. V. Balatsky, {Phys. Rev. Lett.} {\bf 105}, 045302 (2010).
\bibitem{Su10c} J.-J. Su, M. J. Graf, and A. V. Balatsky, preprint arXiv:1007.1937.

\bibitem{electron_glass} A. Amir, Y. Oreg, and Y. Imry, Annu. Rev. Cond. Mat. Phys., to appear (2011), preprint  arXiv:1010. 5767.

\bibitem{bert} L. Berthier and G. Biroli, arXiv: 1011.2578 (2010) 

\bibitem{nelsonGF} D. R. Nelson, {\it Defects and Geometry in Condensed Matter Physics}, (Cambridge University Press, Cambridge, 2002).

\bibitem{granato56} A. Granato and K. L\"ucke, J. Appl. Phys. {\bf 27}, 583 (1956).


\bibitem{lubchenkowolynes} V. Lubchenko and P. G. Wolynes,
{Annu. Rev. Phys. Chem.} {\bf 58}, 235 (2007); 
M. Tarzia and M. A. Moore, {Phys. Rev. E} {\bf 75}, 031502 (2007);
W. Gotze,  {J. Phys.: Condens. Matter} {\bf 11}, A1 (1999);
J. P. Garrahan,  {J. Phys.: Condens. Matter} {\bf 14}, 1571 (2002); 
G. Tarjus, S. A. Kivelson, Z. Nussinov, and P. Viot,  {J Phys.: Condens. Matter} {\bf 17}, R1143 (2005);
T. Park et al., {Phys. Rev. Lett.} {\bf 94}, 017002 (2005).


\bibitem{Balatsky07} A. V. Balatsky, M. J. Graf, Z. Nussinov, and S. A. Trugman, Phys. Rev. B {\bf 75}, 094201 (2007).


\bibitem{memory1}
G. Bertotti, {\it Hysteresis and Magnetism for Physicists, Materials
Scientists, and Engineers} (Academic Press, New York, 1998).
\bibitem{memory2}
L. C. Struick, {\it Physical Aging in Amorphous Polymers and Other Materials} (Elsevier, Amsterdam,1978).
\bibitem{memory3}
A. P. Young (Editor), {\it Spin Glasses and Random Fields}, 
Series on Directions in Condensed Matter Physics, Vol. 12 (World Scientific, Singapore, 1998).
\bibitem{memory4}One example is that of HD-DVD Blu Ray devices that use the optical hysteresis 
of chalcogenide glasses for data storage. 


\bibitem{Esquinazi} P. Esquinazi (Editor), {\it Tunneling Systems in Amorphous and Crystalline Solids} 
(Springer, Heidelberg, 1998).
\bibitem{Enss} C. Enss and S. Huncklinger, {\it Low-Temperature Physics} (Springer, Heidelberg, 2005).


\bibitem{Yoo2009} C.-D. Yoo and A. Dorsey {Phys. Rev. B} {\bf 79}, 100504(R) (2009).

\bibitem{Syshchenko2009} J. Day, O. Syshchenko, and J. Beamish, {Phys. Rev. B} {\bf 79}, 214524 (2009).

\bibitem{Syshchenko2010} O. Syshchenko, J. Day, and J. Beamish, {Phys. Rev. Lett.} {\bf 104}, 195301 (2010).


\bibitem{Iwasa2010} I. Iwasa, {\it Phys. Rev. B} {\bf 81}, 104527 (2010).
\bibitem{Reppy2010} J. D. Reppy, {Phys. Rev. Lett.} {\bf 104}, 255301 (2010).

\bibitem{Hunt09} B. Hunt, E. Pratt, V. Gadagkar, M. Yamashita, A. V. Balatsky, and  J. C. Davis, Science {\bf 324}, 632 (2009).


\bibitem{Clark2008} A. C. Clark, J. D. Maynard, and M. H. W. Chan, Phys. Rev. B {\bf 77}, 184513 (2008).
\bibitem{Maris2010} H. J. Maris and S. Balibar, to appear in J. Low Temp. Phys.


\bibitem{Lin07} X. Lin, A. C. Clark, and M. H. W. Chan, Nature {\bf 449}, 1025 (2007).

\bibitem{Lin09} X. Lin, A. C. Clark, Z. G. Cheng, and M. H. W. Chan, Phys. Rev. Lett. {\bf 102}, 125302 (2009).


\bibitem{Grigorev07a} V. N. Grigor'ev, V. A. Maidanov, V. Yu. Rubanskii, S.P. Rubets, E. Ya. Rudavskii, A. S. Rybalko, Ye. V. Syrnikov, and V. A. Tikhii, Phys. Rev. B {\bf 76}, 224524 (2007).

\bibitem{Grigorev07b} V. N. Grigor'ev and Ye. O. Vekhov, J. Low Temp. Phys. {\bf 149}, 41 (2007).

\bibitem{Rudavskii10} E. Y. Rudavskii, V. N. Grigor'ev, A. A. Lisunov, V. A. Maidanov, V. Y. Rubanskii, S. P. Rubets, A. S. Rybalko, and V. A. Tikhii, J. Low Temp. Phys. {\bf 158}, 578 (2010).


\bibitem{Su10a} J.-J. Su, M. J. Graf, and A. V. Balatsky, J. Low Temp. Phys. {\bf 159}, 431 (2010).


\bibitem{Aoki07} Y. Aoki, J. C. Graves, and H. Kojima, Phys. Rev. Lett. {\bf 99}, 015301 (2007).

\bibitem{Aoki08} Y. Aoki, J. C. Graves, and H. Kojima, J. Low Temp. Phys. {\bf 150}, 252 (2008).


\bibitem{Andreev07} A. F. Andreev, JETP Lett. {\bf 85}, 585 (2007).
\bibitem{Andreev09} A. F. Andreev, JETP {\bf 109}, 103 (2009).
\bibitem{Korshunov09} S. E. Korshunov, JETP Lett. {\bf 90}, 156 (2009).


\end{thebibliography}
\end{document}